\documentclass[twocolumn,english,notitlepage,superscriptaddress,nofootinbib]{revtex4-1}
\usepackage[T1]{fontenc}
\usepackage[latin9]{inputenc}
\usepackage{mathtools}
\usepackage{bm}
\usepackage{amsmath}
\usepackage{amsthm}
\usepackage{amssymb}
\usepackage{graphicx}
\usepackage{wasysym}

\makeatletter
\usepackage{comment}
\usepackage{color}
\usepackage{xcolor}
\usepackage{amsthm}

\usepackage{babel}

\newcommand{\affone}{University College London, Gower Street, WC1E 6BT London, United Kingdom.}
\newcommand{\afffive}{School of Physics and Astronomy, University of Glasgow, Glasgow, G12 8QQ, United Kingdom.}
\newcommand{\afftwo}{University of Groningen
PO Box 72, 9700 Groningen, The Netherlands.}

\newcommand{\affthree}{Van Swinderen Institute, University of Groningen, 9747 AG Groningen, The Netherlands.}
\newcommand{\afffour}{QOLS, Blackett Laboratory, Imperial College London, SW7 2AZ London, United Kingdom.}

\makeatother

\usepackage{babel}
\begin{document}
\title{Relative Acceleration Noise Mitigation for Nanocrystal Matter-wave
Interferometry: Application to Entangling Masses via Quantum Gravity}
\author{Marko Toro\v{s}}
\affiliation{\affone}
\affiliation{\afffive}
\author{Thomas W. van de Kamp}
\affiliation{\afftwo}
\author{Ryan J. Marshman}
\affiliation{\affone}
\author{M. S. Kim}
\affiliation{\afffour}
\author{Anupam Mazumdar}
\affiliation{\afftwo}
\affiliation{\affthree}
\author{Sougato Bose}
\email{s.bose@ucl.ac.uk}

\affiliation{\affone}
\begin{abstract}
Matter wave interferometers with large momentum transfers, irrespective
of specific implementations, will face a universal dephasing due to
relative accelerations between the interferometric mass and the associated
apparatus. Here we propose a solution that works even without actively
tracking the relative accelerations: putting both the interfering
mass and its associated apparatus in a freely falling capsule, so
that the strongest inertial noise components vanish due to the equivalence
principle. In this setting, we investigate two of the most important
remaining noise sources: (a) the non-inertial jitter of the experimental
setup and (b) the gravity-gradient noise. We show that the former
can be reduced below desired values by appropriate pressures and temperatures,
while the latter can be fully mitigated in a controlled environment.
We finally apply the analysis to a recent proposal for testing the
quantum nature of gravity {[}S. Bose et. al. Phys. Rev. Lett 119,
240401 (2017){]} through the entanglement of two masses undergoing
interferometry. We show that the relevant entanglement witnessing
is feasible with achievable levels of relative acceleration noise. 
\end{abstract}
\maketitle
The two pillars of modern physics, Quantum mechanics and General relativity,
are expected to be eventually combined into the elusive theory of
Quantum Gravity (QG)~\citep{oriti2009approaches,kiefer2006quantum,penrose1996gravity}.
However, whilst separately the two theories are well tested, the former
in the regime of large masses and distances and the latter in the
microscopic regime, no experiments has been able to probe them simultaneously~\citep{hossenfelder2017experimental}.
To facilitate this formidable task one promising approach is the development
of low energy (infrared) QG phenomenology which could eventually,
upon experimental realization, lead to critical experimental hints.
Of course, gravity has been extensively probed in the domain of quantum
field theory in ``classical'' curved spacetime~\citep{brunetti2015advances,wald1994quantum}.
There the \emph{source} of the gravitational field is classical and
the \emph{probe} is quantum mechanical. The most notable result is
given by the Colella-Overhauser-Werner experiment~\citep{colella1975observation},
which has over the years lead to several important matter-wave interferometers~\citep{nesvizhevsky2002quantum,fixler2007atom,rauch2015neutron}
as well as to more recent developments in photon interferometry~\citep{bertocchi2006single,fink2017experimental,restuccia2019photon}.

To reveal quantum features of the gravitational field one promising
approach is to prepare a nonclassical state of a massive system, resulting
in \emph{a quantum source }of the gravitational field. Specific proposals
have been devised to witness the entanglement between two masses mediated
through a gravitational field ~\citep{bose2017spin,marletto2017gravitationally}.
As a {\em classical mediator cannot entangle} two masses \citep{horodecki2009quantum},
gravity, the mediator of the above entanglement, must be quantum \citep{bose2017spin,belenchia2018quantum,marshman2020locality}.
This seems to be currently the only conclusive way to witness the
fundamentally quantum nature of gravity in the laboratory. Each mass
is placed in a superposition of two positions, which can be rephrased
in a suggestive way by employing a general relativistic viewpoint
-- it is a superposition of spacetime geometries~\citep{christodoulou2019possibility}.

An important question is the level of ambient noise under which interference
or entanglement can be detected. This has been estimated under generic
amounts of decoherence \citep{nguyen2020entanglement,chevalier2020witnessing}
and mitigating pressures and temperatures have been estimated for
gas collisions and black-body sources \citep{bose2017spin,van2020quantum}.
However, {\em any} large mass matterwave experiment, which requires
large momentum transfer to achieve sufficient wavefunction splitting
is acutely susceptible to acceleration noise \citep{bose2018matter,pedernales2019motional}.
Recently, it has been claimed that this type of noise acutely affects
the witnessing of entanglement~\citep{Andre} in the quantum nature
of gravity experiment \citep{bose2017spin}.

Here we propose how the two universal dephasing channels -- non-inertial
jitter (i.e., residual acceleration noise) and gravity gradient noise
(GGN) -- that will limit \emph{any} large momentum transfer matterwave
interferometry experiment, can be mitigated. Large momentum transfer
will always be required for scaling matter wave interferometry to
large masses, for which there is wide motivation: not only the quantum
nature of gravity as mentioned above, but also quantum sensing \citep{marshman2020mesoscopic,qvarfort2018gravimetry,armata2017quantum}
and testing the ultimate limits of quantum mechanics \citep{Bassi}.
Both effects induce random relative accelerations between the interfering
paths as well as with the experimental apparatus (the control fields/beam
splitters that create the superposition as well as the measuring devices),
resulting in a loss of visibility.

We obtain simple formulae to describe the loss of coherence due to
non-inertial jitter, induced by gas collisions and photon scattering
on the experimental container, and gravity gradients, induced by external
masses and the intrinsic finite size of matterwave systems. In particular,
the derived expressions depend only on generic properties of any matterwave
experiment and are \emph{independent} on the specific mechanism/protocols
to prepare and recombine the superpositions. In addition, we estimate
the effects for the interferometric setup from \citep{bose2017spin}
and show that this experiment can be made insensitive to the above
two universal noise effects by controlling the environment (see Figs.~\ref{fig:boxsize} and \ref{fig:ggn}). The non-inertial
jitter of the experimental apparatus (in particular, uncontrollable
motion of the magnets) gives rise to an acceleration noise which has
to be kept below $\sim\text{fm\,\ensuremath{\text{s}^{-2}}}/\text{\ensuremath{\sqrt{\text{Hz}}}}$
~\citep{Andre} -- we find such a value can be achieved by placing
the experiment in a vacuum chamber with pressure $\sim10^{-6}$ Pa
or lower (one can have further technical noises, e.g. vibrations,
rotational noise, charge noise, which have to be controlled to the
same degree, but similarly do not pose a fundamental limitation).
Gravity-gradient noise can be on the other hand mitigated by limiting
access to the immediate vicinity of the experiment to massive moving
bodies (e.g. to $\sim5$m for humans, $\sim10$m for cars, and $\sim60$m
for planes).

This work is organized in the following way. We discuss how to derive
the Lagrangian appropriate for interferometric protocols starting
from Fermi-Normal coordinates (Sec.~\ref{sec:Non-inertial-Reference-frames}).
We then discuss phase accumulation in single particle interferometry
experiments~\citep{marshman2020mesoscopic}, in particular, focusing
on non-inertial jitter and the resulting loss of visibility. We then
discuss the gravity gradient noise (GGN) due to finite size effect
of the capsule and outline its mitigation (Sec.~\ref{sec:noises}).
We finally apply the results to the experimental setup to detect the
quantum nature of gravity~\citep{bose2017spin} where we consider
an improved scheme to reduce the Casimir-Polder interaction~\citep{van2020quantum}.
Here we give quantitative estimates for the entanglement witness under
feasible mitigations of the above noise sources using a recently proposed
improved entanglement witness~\citep{chevalier2020witnessing} (Sec.~\ref{sec:Spin-entanglement-witnesses}).

\begin{figure}[t]
\includegraphics[width=1\columnwidth]{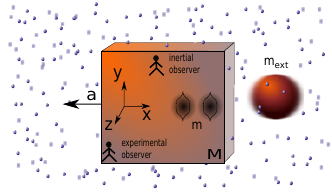}\centering

\caption{Conceptual scheme of the experiment as seen by a distant observer.
Here we focus on the horizontal $x$-motion where the objects of mass
$m$ are placed in a spatial superposition. Both the system (here
depicted as two adjacent interferometers) and the experimental apparatus
(here illustrated as a box of mass $M$) follow approximately geodetic
motion. The deviation from ideal geodetic motion is due to gas collisions
and photon scattering (here we have illustrated only dust particles
outside the experimental container). To describe the experiment we
consider an ideal free-falling observer and an observer attached to
the experimental container. For the ideal free falling observer also
the experimental-box becomes a dynamical degree of freedom (to account
for its motion about the geodesic), while the observer attached to
the experimental container will describe it using an accelerated reference
frame with a time-dependent acceleration $a$. In addition, any external
mass $m_{\text{ext}}$ will generate a small gravity-gradient over
the finite extension of the experiment. On the other hand, the uniform
potential generated by the same mass $m_{\text{ext}}$ vanishes due
to the equivalence principle -- for the experimental observer both
the system and the apparatus fall at the same rate towards any external
mass.}
\label{fig:conceptual} 
\end{figure}

\section{Reference frames and matter-waves\label{sec:Non-inertial-Reference-frames}}

A convenient coordinate system to describe matter-wave experiments
is the one where the experimental equipment remains stationary (see
Fig.~1). Assuming that the experimental apparatus is attached to
a container -- forming an experimental box -- one can consider the
motion of its center-of-mass and construct the associated time-like
curve in spacetime. Here we are assuming that the mass of the system,
$m$, is much smaller then the total mass of the experimental box,
$M$, such that the effect of the former on the latter can be neglected.
To describe such nearly-local interferometric experiments we first
construct Fermi-Normal coordinates (FNC). In particular, the FNC metric
is given by~\citep{misner1973gravitation,poisson2011motion}:

\begin{alignat}{1}
ds^{2} & =g_{tt}c^{2}dt^{2}+2g_{tb}cdtdx^{b}+g_{bc}dx^{b}dx^{c},\label{eq:fnc}\\
g_{tt} & =-[(1+a_{b}x^{b})^{2}+R_{0c0d}x^{c}x^{d}],\\
g_{tb} & =-\frac{2}{3}R_{0cbd}x^{c}x^{d},\\
g_{bc} & =\delta_{bc}-\frac{1}{3}R_{bcde}x^{d}x^{e},
\end{alignat}
where we have omitted cubic displacements $\mathcal{O}(x^{3})$ from
the reference time-like curve, and $\bm{a}=(a_{1},a_{2},a_{3})$ is
the acceleration of the observer. The curvature effects are encoded
in the Riemann tensor $R$ which can be estimated from the background
stress-energy tensor. Here we are also implicitly assuming that the
reference frame is not rotating as we have restricted the discussion
only to linear accelerations. The FNC construction is typically applied
to investigate classical Earth-Bound experiments as well experiments
in free-fall~\citep{will2006confrontation}.

For non-relativistic matter-wave experiments we can make further approximations.
In particular, for slowly moving matter only the $g_{tt}$ term will
be important, i.e. when expanding the dynamics to order $\mathcal{O}(c^{-1})$.
We thus approximate the metric in Eq.~(\ref{eq:fnc}) to:

\begin{equation}
ds^{2}=-((1+a_{b}x^{b})^{2}+R_{0c0d}x^{c}x^{d})^{2}c^{2}dt^{2}+\delta_{bc}dx^{b}dx^{c}.\label{eq:metric}
\end{equation}
In many cases the curvature effects are negligible, i.e. we can further
neglect the Riemann tensor term $\sim R_{0c0d}$ in Eq.~(\ref{eq:metric}),
resulting in the Rindler metric. We however keep the term $R_{0c0d}x^{c}x^{d}$,
which corresponds to Newtonian spacetime curvature, i.e. the gravity-gradient
term, which can result in relative accelerations between the mass
and the measuring apparatus if they are finitely spatially separated
in the laboratory. We keep this in order to examine the influence
of GGN, but as we will show, it can be mitigated for all reasonable
unknown masses that cannot be tracked during the experiment.

In any case, we can readily write down the Lagrangian of a point particle:

\begin{equation}
L=-mc^{2}\sqrt{-g_{\mu\nu}\frac{dx^{\mu}}{cdt}\frac{dx^{\nu}}{cdt}},\label{eq:lagrangiandef}
\end{equation}
where $x^{\mu}=(ct,\bm{x}$) are the FNC coordinates. Since we are
primarily interested in the motion along the horizontal direction,
i.e. the axis of the spatial superposition, we will in the following
omit the coordinates $x_{2}$, $x_{3}$ and relabel $x_{1}$ ($a_{1}$)
as $x$ ($a$). Using the metric in Eq.~(\ref{eq:metric}) and the
Lagrangian in Eq.~(\ref{eq:lagrangiandef}) we then readily obtain

\begin{equation}
L=\frac{1}{2}mv^{2}-max-\frac{1}{2}m\omega_{\text{gg}}^{2}x^{2},\label{eq:lagrangian}
\end{equation}
where we have omitted the constant term $mc^{2}$, and we have introduced
$\omega_{\text{gg}}^{2}=R_{0101}c^{2}$. The harmonic frequency $\omega_{\text{gg}}$
is associated with the Newtonian \emph{gravity-gradient} potential
due to finite size of the experiment: for an attractive one it is
real-valued, but for a repulsive one it becomes imaginary. Physically
this corresponds to tidal forces that are compressing or stretching
a body, respectively.

We now concentrate on the setting of Fig.~1. The whole experiment
is enclosed in a free fall laboratory (which we also interchangeably
call the capsule or box). The Lagrangian we have obtained in Eq.~(\ref{eq:lagrangian})
describes the motion of the system from the viewpoint of the non-inertial
laboratory observer (i.e., comoving with the experimental box). It
is important to note that the acceleration $a$ can only result from
electromagnetic interactions but not through the gravitational one,
e.g. dust particle or photons hitting the experimental-box. Importantly,
a laboratory interacting only gravitationally with external masses
would still result in free fall with vanishing acceleration, i.e.
$a=0$. Indeed, from the viewpoint of a distant inertial observer
both the experimental-box as well as the system would be accelerating
towards the external mass with the same acceleration, $Gm_{\text{ext}}/R^{2}$,
where $m_{\text{ext}}$ is the mass of the external object, and $R$
is the distance between the external object and the center of the
experimental-box. On the other hand, gravity-gradient potentials here
parameterized by $\omega_{\text{gg}}$, cannot be eliminated by simple
change of coordinates, as quantum-mechanical systems are always of
finite extension due to their wave-nature.

In summary, one can repeat the FNC construction for different observers,
following different time-like curves. In this section we have already
discussed three different observers, each of which has a different
coordinate system: an ideal free-falling observer following a geodesic,
the approximately free-falling observer following the time-like curve
of the experimental box, and the distant inertial observer fixed with
respect to the stars. While the above construction was based on the
general relativistic formalism, the same non-relativistic results
can be obtained directly using extended Galilean transformations.
Importantly, non-inertial effects can be seen as relative motion between
the experimental box and the system, the former following a non-geodesic
time-like curve while the latter on a geodesic (when in perfect isolation).
On top of this, each interaction of a gas particle or a photon with
the system will induce non-geodesic motion of the latter: this gives
rise to the decoherence already considered in~\citep{bose2017spin,van2020quantum}.
On the other hand, gas and photon collisions with the experimental
box provides a second mechanism for the loss of visibility: we will
refer to it as \emph{non-inertial jitter} (sometimes labeled as residual
acceleration noise). However, there is an important difference between
the two: unlike decoherence, the loss of visibility stemming from
non-inertial jitter can, at least in principle, be completely canceled
by a control experiment. Indeed non-inertial jitter, as well as any
other classical deterministic noise, can be measured using a second
system, and addressed either by actively recalibrating the experimental
apparatus in real-time or passively in post-analysis. As we will see
in the next section non-inertial jitter is a technical challenge,
but does not present a fundamental limitation for interferometry with
large masses.

\section{Non-inertial jitter and gravity-gradient noise\label{sec:noises}}

\begin{figure}[t]
\includegraphics[width=0.75\columnwidth]{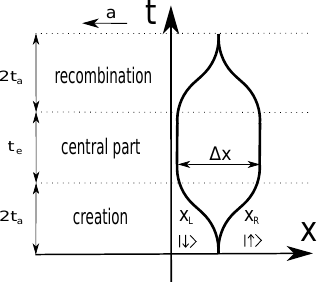}\centering

\caption{Paths for a single particle interferometry experiment as described
from the viewpoint of an observer stationary with the experimental-box
-- we have indicates by $a$ the residual acceleration which arises
due to the collisions of the experimental-box with gas particles (see
Fig.~\ref{fig:conceptual}). The paths are predominantly determined
by the magnetic-field gradient forces (and hence still symmetric),
while the phases can have also unknown random contributions from non-inertial
and gravity-gradient terms. Such phases can induce a dephasing channel
when there is \emph{momentum transfer} between the system and experimental
apparatus (i.e., when there is \emph{relative} motion between the
system and experimental apparatus and the two are coupled). The inteferometric
loop has three parts: (i) creation of superposition, (i) central part
when the system is completely decoupled from the experimental apparatus,
and (iii) recombination of the superposition. Interference and dephasing
can be discussed only when the full interferometric loop is taken
into account (see text).}
\label{fig:scheme2} 
\end{figure}

In this section we consider a single interferometer for a mass $m$
with two internal states $\mathtt{s}_{j}$ ($j=L,R$) -- we create
and control the superposition size by using state dependent forces~\citep{bose2017spin}.
In particular, we first create a spatial superposition, maintain it
in a fixed size $\Delta x$ for a fixed interval of time, and then
recombine it, as shown in Fig.\ref{fig:scheme2}; at the end we measure
the resulting accumulated phase difference. We describe the two paths
of the superposition using the semi-classical approach \citep{storey1994feynman}.
We consider the Lagrangian obtained in Eq.~(\ref{eq:lagrangian})
and add the interaction for controlling the superposition size. Specifically,
for the two paths we have the following Lagrangian:

\begin{equation}
L_{j}=\frac{1}{2}mv_{j}^{2}-ma(t)x_{j}-m\lambda_{j}(t)x_{j}-\frac{1}{2}m\omega_{\text{gg}}^{2}(t)x_{j}^{2},\label{eq:lagrangians-1}
\end{equation}
where $j=L,R$ denotes the left or right path, $x_{j}$ the particle
position, $\lambda_{j}(t)=\frac{f_{\text{m}}}{m}\mathtt{s}_{j}$ is
a state dependent acceleration generated from a force of amplitude
$f_{\text{m}}$ (the internal state labels can acquire values $\mathtt{s}_{j}=\pm1$
during the creation of the superposition and its recombination, while
during the period the superposition is held constant, it is set to
$\mathtt{s}_{j}=0$ (see Fig.~\ref{fig:scheme2})), and $a(t)$ is
the time-dependent acceleration as described by the non-inertial observer
attached to the experimental box (here we are using the term non-inertial
as the box is subject to nonintertial jitter). In Ref.\citep{bose2017spin},
a specific realization of the state dependent force was suggested,
where $\mathtt{s}_{j}$ corresponded to NV centre spin states in a
diamond nanocrystal, and the state dependent force was generated by
a magnetic field gradient $\frac{\partial B}{\partial x}$ through
\begin{equation}
f_{\text{m}}=g_{NV}\mu_{B}\frac{\partial B}{\partial x},\label{eq:fmag}
\end{equation}
where $g_{\text{NV}}$ is the electronic $g$-factor, $\mu_{B}$ is
the Bohr magneton, $B$ is the component of the magnetic field along
$x$. However, here we are going to refrain from the details of the
properties of the crystal and the source of the magnetic field gradient~\citep{scully1989spin}.
Instead, we are going to focus on those relative acceleration noise
sources which would be present in any realization of matter wave interferometry
through generic internal state dependent forces as modeled in Eq.(\ref{eq:lagrangians-1}).
However, for simplicity of presentation we are going to refer to the
internal states as spins.

The \emph{trajectories} for the \emph{two states} associated to the
different initial position and spin are determined by the simple equation

\begin{equation}
\ddot{x}_{j}(t)=\lambda_{j}(t).\label{eq:dynamics}
\end{equation}
Here we have omitted the contribution from gravity gradients of unknown
external masses, i.e. $\omega_{\text{gg}}^{2}=0$, as we are primarily
interested in the trajectories (while the effect of gravity-gradient
terms from known sources on the trajectory can be readily taken into
account in the analysis and are thus also omitted here). Similarly,
unknown sources of $a(t)$ in a controlled environment will be small
and can be neglected, while known sources, such as due to the motion
of the experimental apparatus, can be fully taken into the analysis.
However, the same argument we have applied for trajectories does not
apply to the \emph{accumulated phase difference} where already tiny
non-inertial and gravity-gradient contributions could rotate it by
a substantial fraction of $2\pi$. This will be discussed in detail
below. The trajectories are thus determined by 
\begin{equation}
x_{j}=\int_{0}^{t}\left[\int_{0}^{u'}\lambda_{j}(u)du\right]du',\label{eq:free_fall}
\end{equation}
where we have assumed $x_{j}(0)=0$ and $\dot{x}_{j}(0)=0$.

For a single particle we can generate two distinct paths, i.e. $j=L$
and $j=R$ (left and right paths, respectively), by considering opposite
spins , i.e. $\mathtt{s}_{L}=-\mathtt{s}_{R}$, such that the magnetic
forces are opposite: $\lambda_{L}=-\lambda_{R}$. The condition to
close the loop at time $t=t_{f}$ is given by requiring: 
\begin{equation}
x_{L}(t_{f})=x_{R}(t_{f}).
\end{equation}
As the state dependent force depends linearly on the spin of the particle,
i.e. $\propto\mathtt{s}_{j}$, this give a condition on the time-dependence
of the spin values

\begin{equation}
\int_{0}^{t}\left[\int_{0}^{u'}\mathtt{s}_{j}(u)du\right]du'=0,\label{eq:simplified_condition}
\end{equation}
i.e. the condition to \emph{close} the interferometric loop. For example,
Eq.~(\ref{eq:simplified_condition}) can be fulfilled by controlling
the spins as follows:

\begin{equation}
\mathtt{s}_{L}(t)=\begin{cases}
-1, & 0<t<t_{a},\\
+1, & t_{a}<t<2t_{a},\\
0, & 2t_{a}<t<2t_{a}+t_{e},\\
+1, & 2t_{a}+t_{e}<t<3t_{a}+t_{e},\\
-1, & 3t_{a}+t_{e}<t<4t_{a}+t_{e},
\end{cases},\label{eq:spins}
\end{equation}
with the opposite values for $\mathtt{s}_{R}(t)$. The total experimental
time is given by $t_{f}=4t_{a}+t_{e}$, where we will refer to $t_{a}$
($t_{e}$) as the acceleration (free-fall) time interval. Even if
this condition is not exactly met experimentally, as long as the final
states are approximately equal, i.e. with nearly overlapping wavepackets,
one will not have substantial loss of visibility: if the spread of
the wavepackets is $\sigma$, one requires $\vert x_{L}(t_{f})-x_{R}(t_{f})\vert\ll\sigma$.
Note that a random acceleration $a(t)$ does not affect at all the
condition $x_{L}(t_{f})=x_{R}(t_{f})$: both paths are subject to
exactly the same random acceleration $a(t)$ and the loop thus remains
perfectly closed. Only the random fluctuating gravity-gradient term
can affect the closed loop condition when sufficient asymmetry is
present in the problem. Indeed, this is Stern-Gerlach interferometry,
which has recently been implemented with atoms \citep{Folman2013,Folman2018,folman2019}
and suggested for large masses \citep{scala2013matter,PhysRevLett.117.143003}.
We should note, however, this criterion can be difficult to meet,
and it is eased by cooling the masses to the ground state initially
in a trap, which has already been achieved \citep{delic2020cooling}.

In the next two sections we will estimate two dephasing channels using
the method which we now sketch. The accumulated phase difference for
the interferometric loop is given by

\begin{equation}
\Delta\phi=\phi_{R}-\phi_{L},\label{eq:difference}
\end{equation}
where the accumulated phase over each path $j=L,R$ is given by

\begin{equation}
\phi_{j}=\frac{1}{\hbar}\int_{0}^{t_{f}}dtL_{j}(t),\label{eq:phij}
\end{equation}
and $t_{f}$ is the time of interferometric experiment. The measured
phase can be written as $\Delta\phi=\Phi_{\text{eff}}+\delta\phi$,
where $\Phi_{\text{eff}}$ would be the phase in absence of dephasing
or decoherence channels, and $\delta\phi$ is the fluctuating contribution
due to noise sources. We will investigate the decay of coherences,
$\Gamma\sim\mathbb{E}[\delta\phi^{2}]$, arising from non-intertial
jitter and GGN, $a(t)$ and $\omega_{\text{gg}}^{2}(t)$, respectively,
where $\mathbb{E}[\,\cdot\,]$ denotes the average over different
noise realizations. In particular, the condition to witness interference
(as well as entanglement) can be cast in the form:

\begin{equation}
\Phi_{\text{eff}}>\Gamma_{\text{jitter}}+\Gamma_{\text{gg}},\label{eq:requirement}
\end{equation}
where $\Gamma_{\text{jitter}}$ and $\Gamma_{\text{gg}}$ denote the
decay of coherences due to non-inertial jitter and GGN, respectively
(any other channel for the loss of visibility will appear on the right-hand
side as and additional contribution).

\subsection{Non-inertial noise/residual acceleration noise \label{sec:Non-inertial noise}}

We want to calculate the accumulated phase difference arising from
the non-inertial jitter of the experimental box (sometimes labeled
as residual acceleration noise in the literature) and estimate its
effect on the interferometric visibility\footnote{The gravity-gradient terms, which will be discussed in the next subsection,
are here set to zero, i .e. $\omega_{\text{gg}}=0$.}. Exploiting the Lagrangian in Eq.~(\ref{eq:lagrangians-1}) and
the trajectories given by

\begin{equation}
x_{j}=\int_{0}^{t}\left[\int_{0}^{u'}(\lambda_{j}(u)+a(u))du\right]du',\label{eq:free_fall2}
\end{equation}
we eventually find a simple expression

\begin{equation}
\Delta\phi=\frac{2m}{\hbar}\int_{0}^{t_{f}}dt\lambda(t)X(t),\label{eq:deltaphi}
\end{equation}
where we have defined

\begin{equation}
X(t)=\int_{0}^{t}\left[\int_{0}^{u'}a(u)du\right]du'.
\end{equation}
We note that from the perspective of the inertial observer $X(t)$
corresponds to the displacement of the experimental box about the
geodesic trajectory generated by gas collision (the analogous effect
generated by photons follows the same analysis).

We can thus readily model the motion of the center-of-mass of the
experimental box as a classical degree of freedom: $X$ ($P$) will
be a classical position (conjugate momentum) observable of the experimental-box.
Specifically, we have the following stochastic differential equations~\citep{bowen2015quantum}:

\begin{alignat}{1}
\dot{X} & =\frac{P}{M},\label{eq:X}\\
\dot{P} & =-\Omega^{2}X-\gamma P+\sqrt{2\gamma Mk_{B}T}P_{\text{in}}.\label{eq:P}
\end{alignat}
The gas-damping coefficient is given by~\citep{cavalleri2009increased}:

\begin{equation}
\gamma=\frac{pl^{2}}{M}(1+\frac{\pi}{8})\left(\frac{32m_{g}}{\pi k_{b}T}\right)^{1/2},\label{eq:gamma}
\end{equation}
where $p$ ($T$) is the gas pressure (temperature), $m_{g}$ is the
mass of a gas molecule, and $l$ is the linear size of a cubic experimental
box. Here we have also included for completeness the harmonic frequency,
$\Omega$, which has to be taken into account, for example, when the
experimental setup is suspended. In the following we however set it
to zero, i.e., $\Omega=0$, as is the case for a free falling setup.
Even if $\Omega$ is nonzero, as long as $\Omega<\omega_{\text{min}}=2\pi t_{\text{exp}}^{-1}$,
where $t_{\text{exp}}$ is the experimental time, we can safely neglect
it. $P_{\text{in}}$ is the classical input noise quantified by:

\begin{equation}
\mathbb{E}[P_{\text{in}}(t)]=0\qquad\mathbb{E}[P_{\text{in}}(t)P_{\text{in}}(t')]=\delta(t-t'),
\end{equation}
where $\mathbb{E}[\,\cdot\,]$ denotes the average over different
noise realizations. To describe the non-inertial jitter of the experimental-box
induced by photons one has to a use a modified Eq.~(\ref{eq:gamma})
-- but will produce only a subleading effect in a controlled environment
-- which we leave for future work~\citep{seberson2020distribution}.

From Eqs.~(\ref{eq:X})-(\ref{eq:P}) we can readily find the power
spectral density (PSD):

\begin{equation}
S_{XX}(\omega)=\frac{4k_{B}T}{M}\frac{\gamma}{(\Omega^{2}-\omega^{2})^{2}+\omega^{2}\gamma^{2}}.\label{eq:Sxx}
\end{equation}
We note that the noise decreases as $1/\omega^{4}$ thus strongly
suppressing high frequency noise. From Eqs.~(\ref{eq:deltaphi})
and (\ref{eq:Sxx}) we can now find the fluctuations of the accumulated
phase:

\begin{equation}
\Gamma_{\text{jitter}}\equiv\mathbb{E}[\Delta\phi^{2}]=\frac{2m^{2}}{\pi\hbar^{2}}\int_{-\infty}^{\infty}d\omega F_{\text{jitter}}(\omega)S_{XX}(\omega),\label{eq:Gamma}
\end{equation}
where we have defined 
\begin{equation}
F_{\text{jitter}}(\omega)=\left[\int_{0}^{t_{f}}dt\int_{0}^{t_{f}}dt'\lambda_{t}\lambda_{t'}e^{i\omega(t-t')}\right].
\end{equation}
Using Eq.(\ref{eq:spins}) the function $F$ is given by

\begin{alignat}{1}
F_{\text{jitter}}(\omega)= & \left(\frac{f_{\text{m}}}{m}\right)^{2}\frac{64\sin^{4}\left(\frac{\omega}{2}t_{a}\right)\sin^{2}\left(\frac{\omega}{2}(2t_{a}+t_{e})\right)}{\omega^{2}}\label{eq:Fexact}
\end{alignat}
which we note is symmetric in $\omega$.

It is instructive to further explore the regime of low damping as
the experiment is expected to be in a controlled environment. In particular,
we consider the case when the damping $\gamma$ is small on the time-scale
of the experiment, i.e. $\omega>\gamma$. We can thus simplify Eq.~(\ref{eq:Sxx})
as:

\begin{equation}
S_{XX}(\omega)\approx\frac{4k_{B}T}{M}\frac{\gamma}{\omega^{4}}.\label{eq:Sxxs}
\end{equation}
We can now use Eqs.~(\ref{eq:Gamma})-(\ref{eq:Sxxs}), to obtain
a simple formula for the phase fluctuations:

\begin{equation}
\Gamma_{\text{jitter}}=\frac{16\gamma k_{B}Tf_{\text{m}}^{2}}{\hbar^{2}M}\left[\frac{23}{15}t_{a}^{5}+t_{a}^{4}t_{e}\right].\label{eq:GammaSimple}
\end{equation}
Let us briefly discuss how to mitigate the phase fluctuations in Eq.~(\ref{eq:GammaSimple}).
Using Eq.~(\ref{eq:gamma}) we first note that Eq.~(\ref{eq:GammaSimple})
has the desired behavior with the pressure, $p$, and temperature,
$T$, of the environment, and can thus be controlled using cryogenics
and vacuum chambers. We can furthermore strongly mitigate the phase
fluctuations by lowering the experimental time $t_{\text{exp}}=4t_{a}+t_{e}$,
for example, by running simultaneously a large number of equal experiments.
In addition, we note that increasing the mass $M$ of the experimental
container also suppress the phase fluctuation, i.e. the jitter of
a heavier experimental box will be smaller with respect to a lighter
one. Specifically, from Eq.~(\ref{eq:gamma}) we find that $\gamma$
scales with the area $l^{2}$, and is inversely proportional to the
mass, i.e. $\sim l^{2}/M$ and thus the overall scaling is $\sim l^{2}/M^{2}$
$\sim1/(\rho^{2}l^{4})$, where $l$ ($\rho$) denote the linear size
(average density) of the experimental box.

On the other hand, it is interesting to note that the phase fluctuations
in Eq.~(\ref{eq:GammaSimple}) are completely independent of the
particle mass $m$. The only dependence of the matterwave state is
through the superposition size $\Delta x$, which can be seen from
the dependency on the state dependent force $\sim f_{\text{m}}$ and
the acceleration/deceleration time interval $\sim t_{a}$, i.e. larger
values will make $\Delta x$ larger. Of course if one wants to generate
the same superposition size for a heavier particle as the one achieved
by a lighter one, $f_{\text{m}}$ would need to be increased by the
ratio of their masses. However, for a fixed $f_{\text{m}}$ the loss
of visibility is completely independent of the mass of the matter-wave
system $m$: this shows that nanoscale and microscale interferometry
presents, as far as non-inertial jitter is concerned, the same level
of experimental challenge as atomic interferometry.

\begin{figure}[t]
\includegraphics[height=6cm]{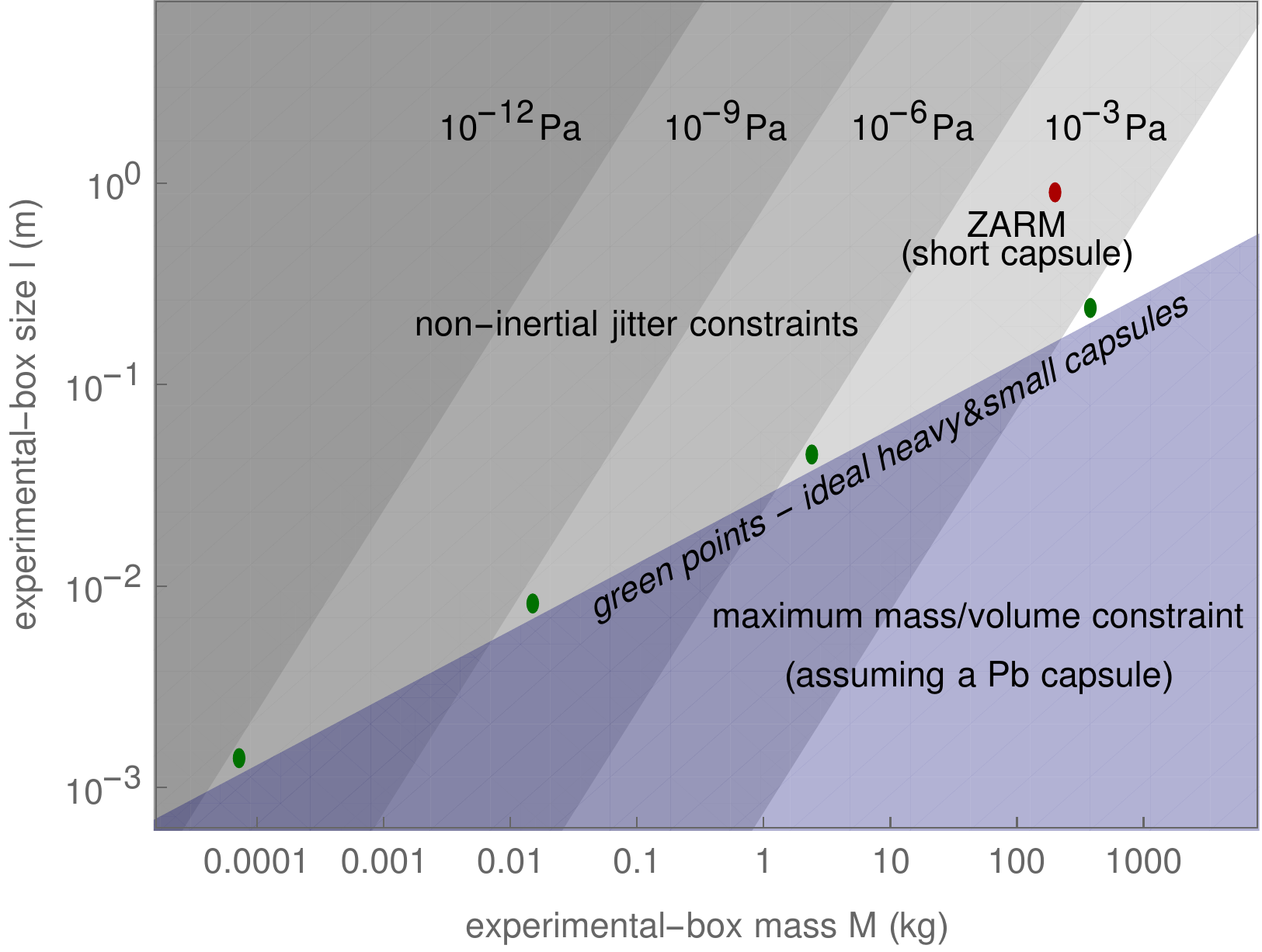}\centering

\caption{Plot of the condition to witness entanglement given by $\Phi_{\text{eff}}>\Gamma_{\text{jitter}}(p,T,l,M)$
where $\Phi_{\text{eff}}$ is the effective entanglement phase, and
$\Gamma_{\text{jitter}}$ is the damping of coherences due to non-inertial
jitter -- the horizontal (vertical) axis denotes the box mass $M$
(box size $l$, i.e., edge length). Using the experimental values
(see Sec.~\ref{sec:Spin-entanglement-witnesses}) we find the effective
entanglement phase $\Phi_{\text{eff}}\sim0.01$ which gives the constraint
$\Gamma_{\text{gg}}\ll0.01$. We consider the outside of the experimental
box to be at room temperature, $T=300\text{K}$, and set the pressure,
$p$, to the following values, $10^{-3}\text{Pa}$, $10^{-6}\text{Pa}$,
$10^{-9}\text{Pa}$, and $10^{-12}\text{Pa}$ -- the corresponding
excluded parameter space is depicted in shades of gray. The pressure
inside the box is set to $\sim10^{-16}\text{Pa}$ and hence its effect
on center-of-mass motion of the experimental box can be neglected.
In addition, we have colored in light blue the region which would
require capsule densities larger than of Lead (Pb) -- the allowed
parameter regime is thus restricted to the upper part of the plot.
We find that non-inertial jitter is successfully suppressed as long
as the pressure, $p$, is low enough for a given box mass/size. Ideally
we would like to have the lightest capsule mass, $M$, which would
allow for simple experimental manipulation -- we have indicated ideal
small and heavy capsules by green points. The limiting cases are given
by a $\sim1.5\text{mm}$ ($\sim25\text{cm}$) size capsule which would
require $10^{-12}$Pa ( $10^{-3}$Pa ) depicted by the green dot in
lower left (upper right) corner. We have also indicated in the plot
a reference point corresponding to the ZARM short capsuale, i.e.,
$l\sim0.9\text{m}$ and $M\sim200\text{kg}$ \citep{fabmbh2008zarm},
well inside the allowed parameter space at $p\sim10^{-6}\text{Pa}$.
Lowering the outside pressure and temperature would even further relax
the constraints on the mass and size of the experimental box.}
\label{fig:boxsize} 
\end{figure}

It is instructive to recast the analysis in terms of an acceleration
noise. Specifically, we consider a simplified model for the center-of-mass
motion of the experimental box:

\begin{alignat}{1}
\dot{X} & =\frac{P}{M},\label{eq:XA}\\
\dot{P} & =MA_{\text{in}},\label{eq:PA}
\end{alignat}
where the input noise is defined by

\begin{equation}
\mathbb{E}[A_{\text{in}}(t)]=0\qquad\mathbb{E}[A_{\text{in}}(t)A_{\text{in}}(t')]=S_{AA}\delta(t-t'),
\end{equation}
$S_{AA}$ is a constant acceleration noise power spectral density,
and $\mathbb{E}[\,\cdot\,]$ denotes the average over different noise
realizations. The more complete model for the center-of-mass motion
of the experimental box discussed above (given in Eqs.~\eqref{eq:X}
and \eqref{eq:P}) can be formally reduced to the simplified model
(given by Eqs.~\eqref{eq:XA} and \eqref{eq:PA}) by making similar
assumptions as used above (see steps from Eq.~\eqref{eq:Sxx} to
Eq.~\eqref{eq:Sxxs}). First, one assumes the mechanical frequency
of the experimental box, $\Omega$, is vanishingly small (such as
in free fall). Second, one assumes that the damping rate, $\gamma$,
is small enough such that the damping term, $-\gamma P$, can be omitted.
Third, the temperature of the environment, $T$, is very large such
that the term $\sqrt{2\gamma Mk_{B}T}P_{\text{in}}$ converges to
the finite force noise term $MA_{\text{in}}$ -- formally, one is
considering an environment in the infinite temperature limit, $T\rightarrow\infty$,
with vanish damping, $\gamma\rightarrow0$, such that the product
$\gamma T$ remains finite. Anyhow, the simplified model gives in
place of Eq.~\eqref{eq:Sxx} the following displacement spectra:

\begin{equation}
S_{XX}(\omega)=\frac{S_{AA}}{\omega^{4}}.\label{eq:SxxA}
\end{equation}
By then comparing Eqs.~\eqref{eq:Sxxs} and \eqref{eq:SxxA} one
can thus extract the following relation:

\begin{equation}
S_{AA}(\omega)\sim\frac{4k_{B}\gamma T}{M}.\label{eq:SxxsA}
\end{equation}
Specifically, Eq.~\eqref{eq:SxxsA} can be used to estimate the acceleration
noise power spectral density from the physical parameters of the problem
(such as the ones discussed in Fig.~\ref{fig:boxsize}). In addition,
we use the relation $\Delta x/2\sim\frac{f_{\text{m}}}{m}t_{a}^{2}$
(see Fig.~\ref{fig:scheme2} and Eqs.~\eqref{eq:free_fall}-\eqref{eq:spins})
to finally rewrite Eq.~\eqref{eq:GammaSimple} as

\begin{equation}
\Gamma_{\text{jitter}}\sim\frac{S_{AA}m^{2}\Delta x^{2}(t_{\text{e}}+\frac{23}{15}t_{\text{a}})}{\hbar^{2}}.\label{eq:GammaSimpleNIJ}
\end{equation}
Non-inertial jitter (i.e., residual acceleration noise) has been recently
discussed in \citep{Andre} -- there they have derived an expression
for the loss of coherences formally matching Eq.~\eqref{eq:GammaSimpleNIJ}
in case $t_{\text{e}}\gg\frac{23}{15}t_{\text{a}}$ (see Eq.~(7)
in \citep{Andre}). The derivation of $\Gamma_{\text{jitter}}$ in
\citep{Andre} was based solely on the central part of the interferometric
loop without considering the preparation and recombination of the
superposition (see Fig.~\ref{fig:scheme2}) which resulted in the
need to apply a somewhat arbitrary frequency filter $\tau^{2}/(1+\omega^{2}\tau^{2})$,
with a free parameter, $1/\tau$, which has been later set to be match
the inverse of the evolution time in the central part, i.e. $\tau\sim t_{\text{e}}$
-- this procedure has yielded an additional factor $1/4$ with respect
to the full calculation we have used above (which considered the full
interferometric loop).

One can take two approaches regarding the value of $S_{AA}$. The
approach taken in \citep{Andre} is to extract its value from existing/proposed
experiments~\citep{selig2010drop,armano2016sub}. However, to devise
tailored-made nanocrystal matter-wave experiments a preferred choice
is to develop an underlying theoretical model which captures the physics
of the residual acceleration noise -- this is the approach taken
in this work. We have shown explicitly that the non-inertial jitter
cannot be induced by unknown external masses\footnote{There is gravitational noise due to external masses, but this is a
higher order effect which arises from gravity gradients and the finite
size of the experiments (see Sec.~\ref{subsec:Gravity-Gradient-Noise}
for more details).} -- in full accordance with the equivalence principle -- but rather
is of electromagnetic origin (dust particles/photons hitting the experimental
apparatus) and can thus be successfully reduced in a controlled environment.
Importantly, using Eqs.~\eqref{eq:SxxsA} and \eqref{eq:GammaSimpleNIJ}
one can design matter-wave interferometry experiments with nano and
micro-size particles -- specifically, one can find the requirements
on the experimental box (size and mass) and on the environment (pressure
and temperature) to successfully perform the experiment\footnote{We have shown that non-inertial jitter (i.e., residual acceleration
noise) is \emph{not} a fundamental limitation of the proposed Quantum-gravity-entangling-of-masses
scheme~\citep{bose2017spin} -- or for this matter for \emph{any}
nano and micro-scaled interferometric scheme -- but rather can be
mitigated by considering a heavy experimental box (sometimes referred
as capsule) in a low pressure environment (see Fig.~\ref{fig:boxsize}).}.

\subsection{Gravity Gradient Noise due to Finite Size Effects\label{subsec:Gravity-Gradient-Noise}}

Gravity gradient noise (GGN), as described by Eq.~\eqref{eq:metric}
will arise from stochastic variations in the curvature which remain
as an external gravitational signal even in a nearly-local experiment
due to its finite size. In place of the Lagrangian in Eq.~\eqref{eq:lagrangians-1}
we now consider:

\begin{equation}
L_{j}=\frac{1}{2}mv_{j}^{2}-m\lambda_{j}(t)x_{j}-\frac{1}{2}m\omega_{\text{gg}}^{2}(t)x_{j}^{2},\label{eq:lagrangians2}
\end{equation}
where we assume $\omega_{\text{gg}}^{2}(t)$ is a multiplicative noise.
Here we are considering only the GGN from movements of untracked external
masses, while any contributions from known masses can be measured
and taken into account in the analysis without any loss of visibility.
Here we omit the linear acceleration term $\sim a(t)$, which models
the non-inertial jitter, and has been already discussed in Sec.~\ref{sec:Non-inertial noise}.

GGN on free test masses has primarily been estimated in the gravitational
wave detection literature \citep{saulson1984terrestrial,hughes1998seismic,thorne1999human,harms2019terrestrial},
from which, instead of $\omega_{\text{gg}}^{2}(t)$, which is the
key quantity relevant for us, it is the random accelerations $a_{\text{rand}}$
of test masses, which is readily available. The calculations of $a_{\text{rand}}$
are based on the cumulative Newtonian effect of environmental mass
movement noises on a free test mass, but quite independent of the
specifics of gravitational wave detectors, so that it is readily usable
in our case. While this acceleration noise $a_{\text{rand}}$ itself
will be completely eliminated in our proposed free-fall laboratory,
it can be used to estimate the noise in $\omega_{\text{gg}}^{2}$,
following Ref.~\citep{visser2018post} to give

\begin{equation}
S_{\omega_{\text{gg}}^{2}\omega_{\text{gg}}^{2}}(\omega)\equiv\frac{1}{\bar{r}^{2}}S_{a_{\text{rand}}a_{\text{rand}}}(\omega)=\frac{\bar{a}^{2}}{\bar{r}^{2}}\frac{1}{(\frac{\omega}{C})^{\alpha}},\label{eq:S2}
\end{equation}
where we have introduced the strength of the local acceleration fluctuations,
$\bar{a}$, a length-scale parameter $\bar{r}$ characterizing the
distance to the GGN sources, and a decay integer $\alpha>1$ which
depends on the type of source. $C=2\pi\times1\text{Hz}$ is a constant
that fixes the correct dimensions. In a more refined analysis one
would need to consider all external masses, and their associated stochastic
motions, which would determine the value of $\frac{\bar{a}^{2}}{\bar{r}^{2}}$
as well as of $\alpha$ in Eq.~\eqref{eq:S2} -- for a fixed value
$\bar{a}$ one can interpret $\bar{r}$ as a characteristic length-scale
of all the GGN sources combined~\citep{saulson1984terrestrial}.

It is instructive to obtain an upper-bound on the noise spectrum $S_{\omega_{\text{gg}}^{2}\omega_{\text{gg}}^{2}}(\omega)$
by considering the smallest possible distance from the experiment
$r_{\text{min}}$ which could contain the bulk of the GGN sources.
The environment around the experiment, i.e. located at $r<r_{\text{min}}$,
can be well controlled by the experimentalists (for example, it could
correspond to the inside of the building in a drop-tower experiment),
and as such will not contribute to GGN. On the hand, any external
mass at distances larger than $r_{\text{min}}$ will give a smaller
contribution to the GGN as it would if its motion reached $r_{\text{min}}$.
In other words, here we will assume that \emph{all} of the GGN sources
reach the outer perimeter of the controlled laboratory environment,
i.e. we will set $\bar{r}=r_{\text{min}}$ in Eq.~\eqref{eq:S2},
which will give an \emph{upper-bound} on the noise -- in practice,
the noise will be significantly smaller, with $\bar{r}\gg r_{\text{min}}$.

We can find the phase fluctuations induced by GGN following a similar
analysis as in Sec.~\eqref{sec:Non-inertial noise}. In first instance
we can assume that the trajectory for $x_{j}(t)$ is given by Eq.~\eqref{eq:free_fall},
solely determined by the magnetic gradients. From Eqs.~\eqref{eq:difference},
\eqref{eq:phij} and \eqref{eq:lagrangians2} we then readily find
the accumulated phase difference:

\begin{equation}
\Delta\phi=\frac{m}{2\hbar}\int_{0}^{t_{f}}dt\,\omega_{\text{gg}}^{2}(t)\left(x_{R}(t)^{2}-x_{L}(t)^{2}\right).\label{eq:dphisim}
\end{equation}
Interestingly, for symmetric paths with respect to the origin of the
coordinate system, i.e. $x_{L}(t)=-x_{R}(t)$, we do not have any
accumulated phase difference, i.e. $\Delta\phi=0$, even for a randomly
fluctuating $\omega_{\text{gg}}^{2}(t)$ -- this is a direct consequence
of the harmonic form of the gravity gradient potential. On the other
hand, if we consider asymmetric paths with respect to the origin of
the coordinate system we will have a non-zero value $\Delta\phi$.

For example, when one considers the dynamics in Eq.~\eqref{eq:dynamics}
with the initial condition $\tilde{x}_{j}(0)=\frac{d}{2}$ and $\dot{\tilde{x}}_{j}(0)=0$
one finds the trajectories given by:

\begin{alignat}{1}
\tilde{x}_{j}(t) & =x_{j}(t)+\frac{d}{2},
\end{alignat}
where $x_{j}(t)$ is the trajectory given Eq.~\eqref{eq:free_fall},
i.e. the trajectory with initial condition $x_{j}(0)=0$ and $\dot{x}_{j}(0)=0$.
Using $\tilde{x}_{j}(t)$ in place of $x_{j}(t)$ in Eq.~\eqref{eq:dphisim},
as well as the property $x_{L}(t)=-x_{R}(t)$, we then find:

\begin{equation}
\Delta\phi=\frac{md}{\hbar}\int_{0}^{t_{f}}dt\omega_{\text{gg}}^{2}(t)x_{R}(t).\label{eq:dphi}
\end{equation}

The gravity-gradient fluctuations can be then obtained from $\mathbb{E}[\Delta\phi^{2}]$
following analogous steps as in Sec.~\ref{sec:Non-inertial noise}
where we discussed non-inertial jitter, and $\mathbb{E}[\,\cdot\,]$
denotes the average over different noise realizations. Specifically,
we eventually find the following gravity-gradient fluctuations:

\begin{equation}
\Gamma_{\text{gg}}\equiv\frac{m^{2}d^{2}}{\pi\hbar^{2}}\int_{\omega_{\text{min}}}^{\infty}S_{\omega_{\text{gg}}^{2}\omega_{\text{gg}}^{2}}(\omega)F_{\text{gg}}(\omega),\label{eq:Gammagg}
\end{equation}
where

\begin{equation}
F_{\text{gg}}(\omega)=\int_{0}^{t_{f}}dt\int_{0}^{t_{f}}dt'e^{i\omega(t-t')}x_{R}(t)x_{R}(t'),\label{eq:Fgg}
\end{equation}
we have introduced a low frequency cutoff $\omega_{\text{min}}=2\pi t_{\text{exp}}^{-1}$,
and $t_{\text{exp}}$ is the experimental time.

Using the trajectories in Eq.~\eqref{eq:dynamics} we can explicitly
evaluate Eq.~\eqref{eq:Fgg}: 
\begin{alignat}{1}
F_{\text{gg}}(\omega)=\frac{f_{\text{m}}^{2}}{m^{2}} & \frac{e^{-i\omega(2t_{a}+t_{e})}}{\omega^{6}}\left(t_{a}^{2}\omega^{2}+\left(-1+e^{it_{a}\omega}\right)^{2}e^{it_{e}\omega}\right)\nonumber \\
 & \left(t_{a}^{2}\omega^{2}e^{i\omega(2t_{a}+t_{e})}+\left(-1+e^{it_{a}\omega}\right)^{2}\right),
\end{alignat}
which we can further approximate as

\begin{equation}
F_{\text{gg}}(\omega)=\frac{f_{\text{m}}^{2}}{m^{2}}\left[t_{a}^{4}(t_{a}+t_{e})^{2}\theta(1-t_{a}\omega)+\frac{t_{a}^{4}}{\omega^{2}}\theta(t_{a}\omega-1)\right],\label{eq:F2}
\end{equation}
where we have smoothed over the fast oscillating terms ($\theta$
is the Heaviside step function, i.e., $\theta(x)=0$ for $x<0$ and
$\theta(x)=1$ for $x>0$).

\begin{figure}[t]
\includegraphics[height=6cm]{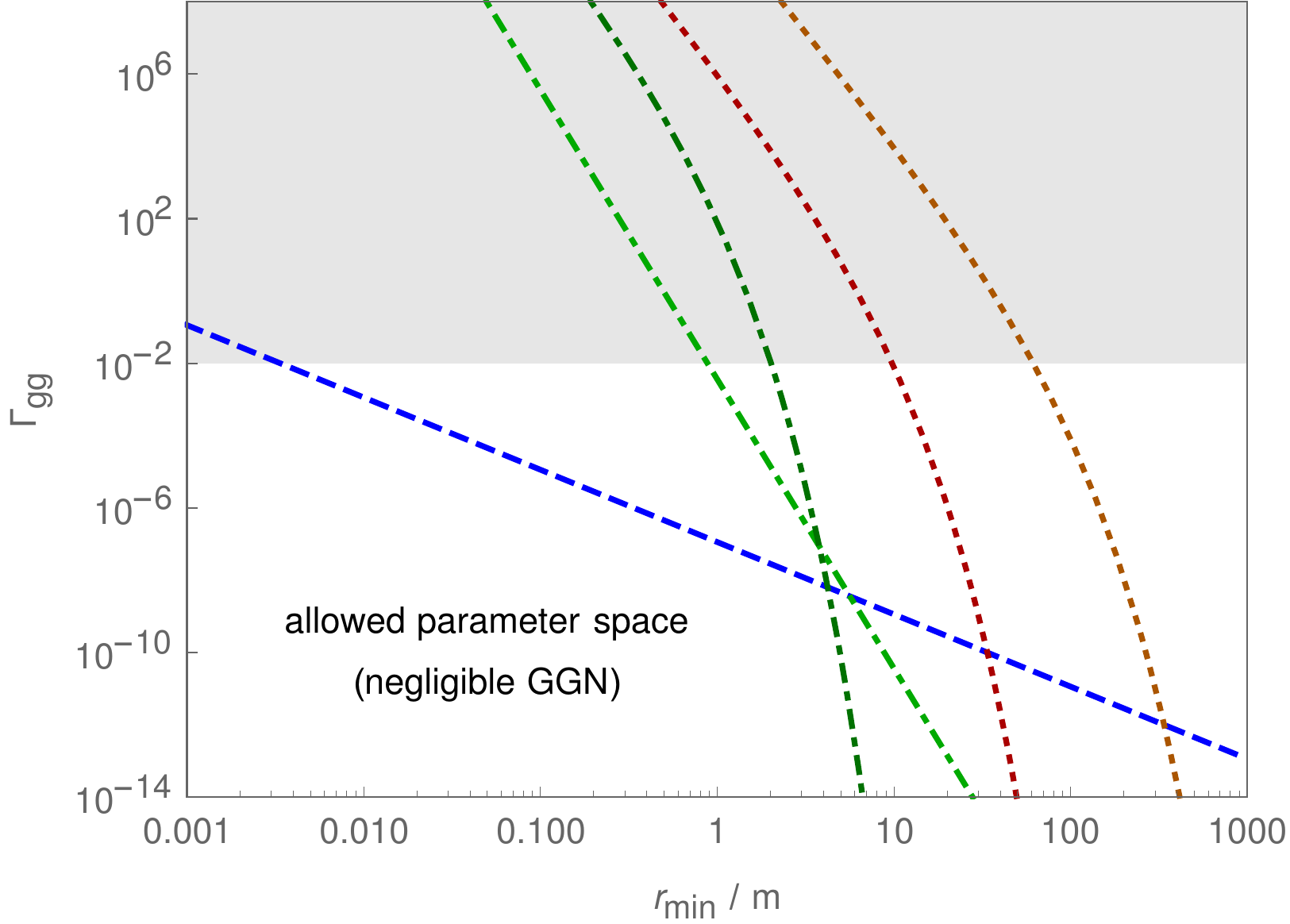}\centering

\caption{Plot of GGN phase fluctuations, $\Gamma_{\text{gg}}$, as a function
of the minimum allowed distance of the respective GGN source from
the experiment, $r_{\text{min}}$, that would still allow the detection
of entanglement. Using the experimental values (see Sec.~\ref{sec:Spin-entanglement-witnesses})
we find the effective entanglement phase $\Phi_{\text{eff}}\sim0.01$
which gives the constraint $\Gamma_{\text{gg}}\ll0.01$ -- the allowed
region is the one colored in white in the lower part of the figure.
We find that seismic and atmospheric activity does not pose a significant
limitation to the experiment (blue dashed line; $r_{\text{min}}\sim0.01\text{m}$).
Similarly, a human is only restricted from walking in the immediate
vicinity of the experiment (light green and dark green dot-dashed
lines for a jerking motion and continuous motion, respectively; $r_{\text{min}}\sim2\text{m}$).
Finally, cars and planes have to be distant by more than $\sim10\text{m}$
and $\sim60\text{m}$, respectively (dark orange and light orange
dotted lines, respectively).}
\label{fig:ggn} 
\end{figure}

Using Eqs.~\eqref{eq:S2}, \eqref{eq:Gammagg}, and \eqref{eq:F2},
and keeping only the dominant term $\sim t_{\text{exp}}^{\alpha-1}$,
we then obtain a simple formula for the gravity-gradient phase fluctuations

\begin{equation}
\Gamma_{\text{gg}}\approx\frac{2\bar{a}^{2}f_{\text{m}}^{2}t_{a}^{4}}{\hbar^{2}}\left(\frac{d}{\bar{r}}\right)^{2}\left[\frac{C^{\alpha}(t_{a}+t_{e})^{2}t_{\text{exp}}^{\alpha-1}}{(2\pi){}^{\alpha}(\alpha-1)}\right].\label{eq:GammaggSimple}
\end{equation}
One first notices that the phase fluctuations in Eq.~\eqref{eq:GammaggSimple}
scale very favorably with the acceleration time, $t_{a}$, as $\sim t_{a}^{4}$
(for the most interesting case $t_{e}\gg t_{a}$), much more favorably
than with the amplitude of the magnetic force, $f_{\text{m}}$, which
scales only as $f_{\text{m}}^{2}$. However the latter parameters
also determine the superposition size $\Delta x\propto$$f_{\text{m}}t_{a}^{2}$,
and for a fixed superposition size there is no benefit of reducing
$t_{a}$ at the cost of increasing $f_{\text{m}}$.

We also note that $\Gamma_{\text{gg}}$ in Eq.~\eqref{eq:GammaggSimple}
does not depend on the mass of the system, $m$, or any other property
of the system. In other words, the phase fluctuations in Eq.~\eqref{eq:GammaggSimple}
are exactly the same for microscale objects as for, say, atoms. Of
course the superposition size, $\Delta x$, scales as $\sim f_{\text{m}}/m$
and thus the superposition size will be smaller for a heavier object
than it would be achieved with a lighter one (compare with phase fluctuations
due to non-inertial jitter in Eq.~\eqref{eq:GammaSimple} which exhibit
a similar behavior).

In this section we have considered the stochastic phase fluctuations
$\Delta\phi$ in Eq.~\eqref{eq:dphi} which then lead to the average
effect $\sim\mathbb{E}[\Delta\phi^{2}]$ in Eq.~\eqref{eq:GammaggSimple}
-- these arise from the trajectories in Eq.~\eqref{eq:free_fall},
solely determined by magnetic forces. In particular, one finds that
$\Delta\phi\propto f_{\text{m}}\omega_{\text{gg}}^{2}(t)$, i.e. the
GGN, $\omega_{\text{gg}}^{2}(t),$ is amplified by the coupling to
the magnetic force, $f_{\text{m}}$. However, there are also tiny
corrections to the trajectories due to non-inertial jitter and due
to the gravity-gradient forces. In particular, in place of Eq.~\eqref{eq:dynamics}
one has a modified dynamics, i.e. $\ddot{x}_{j}(t)=\lambda_{j}(t)+a(t)+\omega_{\text{gg}}^{2}(t)x_{j},$
where the last two terms on the right hand-side are small. Considering
the trajectories perturbed by the noises $a(t)$ and $\omega_{\text{gg}}^{2}(t)$
one finds additional contributions to the phase fluctuation $\Delta\phi$.
For example, from Eqs.~\eqref{eq:difference} and \eqref{eq:phij}
one will find contributions proportional to $\sim a(t)\omega_{\text{gg}}(t)$.
However, the overall phase fluctuation from such terms will be significantly
smaller in comparison to the one in Eq.~\eqref{eq:dphi}, the latter
as discussed amplified by the strong magnetic force, while the former
a product of two weak effects. We leave the full assessment of such
subleading noises for future work.

We finally make a few remarks on the dependency of $\Gamma_{\text{gg}}$
on the parameter $d$. When considering a single interferometer one
can trivially achieve $d=0$, and hence $\Gamma_{\text{gg}}=0$, by
placing the particle initially at the center-of-mass $X$ of the experimental-box.
Indeed, we recall that $d/2$ is by construction the initial displacement
of the particle with respect to $X$. However, in the next section
we will consider two particles in a double interferometric scheme
where we will no longer have the possibility to eliminate the GGN
phase fluctuations simultaneously on both particles (see Fig.~\ref{fig:set-up2}).
In particular, the two particles are placed initially at $\pm d/2$
and one can lo longer avoid the gravity gradient phase fluctuations
in Eq.~\eqref{eq:GammaggSimple} by displacing the two particle about
the center-of-mass of the experimental box. One could, for example,
place one particle at the center-of-mass of the experimental box,
but the other one would be then located, for example, at $d$, which
would result in zero GGN phase fluctuations for the former, but non-zero,
larger ones, for the latter. In addition, the value of $d$ will be
fixed by other experimental requirements and is not vanishingly small.
In short, the GGN phase fluctuations in Eq.~\eqref{eq:GammaggSimple}
can be fully eliminated for a path-symmetric single particle interferometer,
but will have a nonvanishing effect in the double-interferometric
scheme with two particles, which we will consider in the next section.

\section{Testing quantum gravity using a spin entanglement witness\label{sec:Spin-entanglement-witnesses}}

It was shown recently that by imposing a modification to the QGEM-protocol
one can employ a magnetic field gradient two orders of magnitudes
lower than suggested in the original proposal, whilst retaining the
same acceleration/free-fall time intervals~\citep{van2020quantum}.
As such the new experimental setup has a significant effect on noise
reduction. In particular, as shown in Secs. \ref{sec:Non-inertial noise}
and \ref{subsec:Gravity-Gradient-Noise} the non-inertial jitter and
GGN are both proportional to the square of the magnetic field gradient,
hence resulting in a noise reduction by two orders of magnitude.

We will first briefly go over the modified QGEM-protocol and discuss
the effect of technical noises and decoherence effects (Sec.~\ref{sec:setup}).
We will then estimate the phase fluctuations induced by non-inertial
jitter and GGN as well as the how they affect the detectability of
entanglement (Sec.~\ref{subsec:Detectibility-of-entanglement}).
In the latter sections we will be following the modified protocol
as the requirements on the control parameters such as pressure and
temperature will be less demanding, while the methodology of analysis
will resemble the one from the original proposal~\citep{bose2017spin}.
We finally briefly comment on the generality of the noise analysis
and argue that it can be readily adapted also to other matter-wave
experiments (Sec.~\ref{subsec:Detectibility-of-entanglement}).

\subsection{Modified QGEM \label{sec:setup}}

\begin{figure}[h]
\includegraphics[width=1\columnwidth]{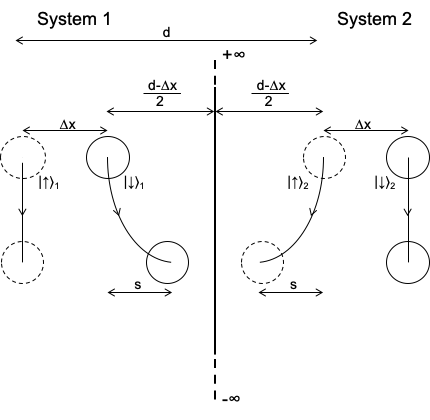} \caption{Modified QGEM protocol. A perfectly conducting plate is placed at
the origin which cancels the Casimir-Polder interaction between the
two masses, allowing for smaller initial separation, $d$. In particular,
one can generate a higher entanglement phase for a given particle
mass. However, we have to take into account the deviation of the particle
trajectories due to the attractive force with the plate -- we denote
the displacement of the inner trajectories towards the plate by $s$
(states associated with spins $\vert\downarrow\rangle_{1}$ and $\vert\uparrow\rangle_{2}$,
where the subscript denotes the particle). \label{fig:set-up2}}
\end{figure}

The modification of the QGEM-protocol recently proposed in ~\citep{van2020quantum}
is illustrated in Fig.~\ref{fig:set-up2}. Specifically, a perfectly
conducting plate is inserted between the two masses -- the plate
is fully reflective to electromagnetic waves, which makes it act like
a \textit{Faraday cage}. The two particles can thus no longer interact
electromagnetically, even at small relative distances, which significantly
relaxes the constraints on their separation -- we want the two particles
to interact only through the weak gravitational interaction, which
is stronger at smaller distances. However, the modification of the
boundary conditions for the electromagnetic field produces an attractive
Casimir force between the plate and each of the two masses -- each
mass moves towards the plate during the free fall by a small displacement,
$s$.

Let us consider in first instance only the unitary part of the dynamics
during the free-fall time, $t_{e}$, while we neglect all other noises
and decoherence channels. One finds that the state at the final time
$t_{f}$ would given by a simple expression 
\begin{alignat}{1}
\vert\Psi(t_{f})\rangle=\frac{1}{2}e^{\text{i}\phi}\bigg[ & {\displaystyle \vert\uparrow\rangle\vert\uparrow\rangle+\vert\downarrow\rangle\vert\downarrow\rangle}\nonumber \\
 & {\displaystyle +e^{\text{i}\Delta\phi_{\uparrow\downarrow}}\vert\uparrow\rangle\vert\downarrow\rangle+e^{\text{i}\Delta\phi_{\downarrow\uparrow}}\vert\downarrow\rangle\vert\uparrow\rangle\bigg]\,,}\label{eq:wavefunction1}
\end{alignat}
where $\vert\,\cdot\,\rangle\vert\,\cdot\,\rangle$ is the joint spin
state for the two particles, and we have omitted the spatial parts
to ease the notation. The accumulated phases are given by: 
\begin{alignat}{1}
\phi & =\frac{Gm^{2}}{\hbar}\int_{0}^{t_{e}}\frac{\text{d}t}{d-s(t)},\\
\Delta\phi_{\uparrow\downarrow} & =\frac{Gm^{2}}{\hbar(d+\Delta x)}t_{e}-\phi,\label{eq:ud}\\
\Delta\phi_{\downarrow\uparrow} & =\frac{Gm^{2}}{\hbar}\int_{0}^{t_{e}}\frac{\text{d}t}{d-\Delta x-2s(t)}-\phi,\label{eq:du}
\end{alignat}
where $G$ is the gravitational constant, and $s(t)$ can be determined
by the Casimir force induced by the plate~\citep{Ford:1998ex}. We
remark that the Casimir interaction will not give rise to the leakage
of ``which-path'' information into the plate, and, thus, will not
be a source of an additional decoherence effect -- its effect is
fully contained in the displacement $s(t)$. There will of course
also be an accumulated phase difference due to the Casimir potential
induced by the plate -- as its value differs on the inner and outer
paths of the individual interferometers. The latter values are however
deterministic and can be fully taken into account, but here we choose
to omit them for simplicity of presentation. In any case, the parameter
which captures the degree of entanglement, namely the effective entanglement
phase, is given by \citep{bose2017spin,marshman2020mesoscopic}: 
\begin{equation}
\Phi_{\text{eff}}=\Delta\phi_{\uparrow\downarrow}+\Delta\phi_{\downarrow\uparrow},\label{eq:phase}
\end{equation}
as can be seen by looking at Eq.~\eqref{eq:wavefunction1}.

We still require that the two interferometric loops remain individually
closed, as otherwise no coherence phenomena can be detected. In particular,
for a single interferometer we require $\vert x_{L}(t_{f})-x_{R}(t_{f})\vert\ll\sigma$,
where $\sigma$ is spread of the wavepackets and $t_{f}=2t_{a}+2\tilde{t}_{a}+t_{e}$
is the total time of the interferometer. The acceleration time interval
to prepare the superposition is $2t_{a}$, but now the time to recombine
the superposition, $2\tilde{t}_{a}$, is longer due the the effect
of the Casimir plate. Specifically, we have 
\begin{equation}
\tilde{t}_{a}=\sqrt{t_{a}^{2}+\frac{s_{\text{max}}}{a_{\text{m}}}},
\end{equation}
where $s_{\text{max}}$ denotes the maximum deviation of the inner
trajectories from free-fall due to the attractive force towards the
plate, $a_{\text{m}}=\frac{f_{\text{m}}}{m}$, and $f_{\text{m}}$
is given in Eq.~\eqref{eq:fmag} (see also Eqs.~\eqref{eq:simplified_condition}
and \eqref{eq:spins}). This latter condition poses a limit on the
minimum separation between the particles and the plate, $d/2$, which
limits the size of displacement induced on by Casimir plate on the
inner paths.

Specifically, we consider two particles with mass $m\sim10^{-15}$
kg placed at the distance $d\sim47\mu$m, a magnetic field gradient
of $\partial_{x}B=10^{4}$ Tm$^{-1}$, the acceleration time $t_{a}\sim0.5\text{s}$,
and the free-fall time $t_{e}\sim1\text{s}$ -- resulting in a superposition
size $\Delta x\sim23\mu$m ( $\mu_{B}\sim9\times10^{-24}\text{J\text{T}}^{-1}$
and $g_{NV}\sim2$). By including the full interferometric loop in
the analysis (including the creation and recombination parts) one
finds $\Phi_{\text{eff}}\sim0.015$ \citep{van2020quantum}.

The full dynamics however contains also non-unitary contributions
as well terms that model technical noises. To describe the final spin
state of the two-particle system we construct a joint density matrix
with basis elements $\vert\uparrow\rangle\vert\uparrow\rangle$, $\vert\uparrow\rangle\vert\downarrow\rangle$,$\vert\downarrow\rangle\vert\uparrow\rangle$,$\vert\downarrow\rangle\vert\downarrow\rangle$
which for brevity we will simply denote by $1,2,3,4$, respectively.
A straightforward calculation eventually gives the density matrix
$\rho$ at time $t_{f}$, which is defined by the following matrix
elements:

\begin{align}
\rho_{11}= & \rho_{22}=\rho_{33}=\rho_{44}=\frac{1}{4},\label{eq:rho11}\\
\rho_{21}= & \rho_{42}^{*}=\frac{1}{4}e^{-\Gamma_{\text{n}}/2-\Gamma_{d}/2+\text{i}\Delta\phi_{\uparrow\downarrow}},\label{eq:rho21}\\
\rho_{31}= & \rho_{43}^{*}=\frac{1}{4}e^{-\Gamma_{\text{n}}/2-\Gamma_{d}/2+\text{i}\Delta\phi_{\downarrow\uparrow}},\\
\rho_{41}= & \frac{1}{4}e^{-2\Gamma_{\text{n}}-\Gamma_{d}},\label{eq:rho41}\\
\rho_{32}= & \frac{1}{4}e^{-\Gamma_{d}/2+\text{i}(\Delta\phi_{\downarrow\uparrow}-\Delta\phi_{\uparrow\downarrow})}.\label{eq:rho32}
\end{align}
The coherences are now damped -- $\Gamma_{\text{n}}$ ($\Gamma_{\text{d}}$)
is the damping of coherences arising from technical noises ( decoherence
effects).

\subsubsection{Technical noises}

The damping from the technical noises is given by: 
\begin{equation}
\Gamma_{\text{n}}=\Gamma_{\text{jitter}}+\Gamma_{\text{gg}},\label{eq:gamman}
\end{equation}
where $\Gamma_{\text{jitter}}$ ($\Gamma_{\text{gg}}$) is given in
Eq.~\eqref{eq:GammaSimple} (Eq.~\eqref{eq:GammaggSimple}). It
is important to note that this damping acts only individually on the
left and right interferometer -- in particular, the degree of entanglement
between the two particles, as quantified by the effective entanglement
phase $\Phi_{\text{eff}}$ in Eq.~\eqref{eq:phase}, remains completely
unaltered by $\gamma_{\text{n}}$. Indeed, the gravitationally induced
entanglement is due to the correlation of the states $\vert\uparrow\rangle\vert\downarrow\rangle$
and $\vert\downarrow\rangle\vert\uparrow\rangle$, which gets fully
encoded in the matrix elements $\rho_{32}$ and $\rho_{23}$. In other
words, technical noises do not change the degree of entanglement,
but can only affect the value of a particular \emph{entanglement witness}.
Thus by carefully measuring the noises, for example using a control
experiment, one could at least in principle fully counteract their
effects thus improving on the interferometric visibility -- we leave
the investigation of such an active scheme for future research.

Let us briefly describe how to derive the damping arising from technical
noises in Eqs.~\eqref{eq:rho21}-\eqref{eq:rho41} -- as discussed
in the previous paragraph these can be derived by considering each
of the two interferometers individually. In a nutshell, a technical
noise will generate a time-dependent, randomly fluctuating phase difference
$\Delta\phi$ between the left and right arm of the interferometer
(see Eq.~\eqref{eq:difference}) -- when considering a large number
of runs of the experiment this will reduce the visibility of the coherences.
In particular, in place of Eq.~\eqref{eq:wavefunction1} we find:
\begin{align}
\vert\Psi(t_{f})\rangle=\frac{1}{2}e^{\text{i}\phi'}\bigg[ & {\displaystyle e^{\text{i}\Delta\phi}\vert\uparrow\rangle\vert\uparrow\rangle+e^{\text{i}\Delta\phi_{\uparrow\downarrow}}\vert\uparrow\rangle\vert\downarrow\rangle}\nonumber \\
+ & {\displaystyle e^{\text{i}\Delta\phi_{\downarrow\uparrow}}\vert\downarrow\rangle\vert\uparrow\rangle+e^{-\text{i}\Delta\phi}\vert\downarrow\rangle\vert\downarrow\rangle\bigg]},\label{eq:wavefunction2}
\end{align}
where $\phi'$ is a common phase. To find the corresponding statistical
operator, averaged over the different runs of the experiment, we calculate
$\hat{\rho}=\mathbb{E}[\vert\Psi(t_{f})\rangle\langle\Psi(t_{f})\vert]$,
where we assume $\mathbb{E}[\Delta\phi]=0$, i.e. a zero-mean fluctuations,
and introduce the variance of the fluctuations $\Gamma\sim\mathbb{E}[\Delta\phi^{2}]$
as discussed in Sec.~\ref{sec:noises}.

\subsubsection{Decoherence}

The total damping of coherences arising from the decoherence channels
is given by~\citep{van2020quantum}: 
\begin{alignat}{1}
\Gamma_{\text{d}}\equiv2 & \sum_{j}\Lambda_{j}\bigg(\frac{46}{15}a_{\text{m}}^{2}\{t_{a}^{5}+\tilde{t}_{a}^{5}\}+4a_{\text{m}}^{2}t_{a}^{4}t_{e}\nonumber \\
 & +\int_{0}^{t_{e}}(4a_{\text{m}}t_{a}^{2}s(t)+s(t)^{2})\text{d}t\bigg)+\Lambda_{\text{air}}t_{f}.\label{eq:fda2}
\end{alignat}
We have three sources of decoherence, namely, the scattering of air
molecules, photon emission and absorption, and photon scattering,
quantified by $\gamma_{\text{air}}\equiv\gamma_{\text{air}}(p_{\text{e}},T_{\text{e}})$,
$\Lambda_{\text{e(a)}}\equiv\Lambda_{\text{e(a)}}(T_{\text{e(i)}})$,
and $\Lambda_{\text{sc}}\equiv\Lambda_{\text{sc}}(T_{\text{e}})$,
respectively, where $T_{\text{e}}$ ($p_{\text{e}}$) is the temperature
(pressure) inside the experimental-box, and $T_{\text{i}}$ is the
internal temperature of particle. The explicit expressions can be
found in \citep{decoherence,PhysRevA.84.052121}:

\begin{alignat}{1}
\gamma_{\text{air}} & =\frac{16\pi n_{V}R^{2}}{3}\sqrt{\frac{2\pi k_{B}T_{ex}}{m_{g}}}\,,\\
\Lambda_{\text{sc}} & =8!\zeta(9)\frac{8cR^{6}}{9\pi}\left(\frac{k_{B}T_{ex}}{\hbar c}\right)^{9}\text{Re}\left(\frac{\epsilon-1}{\epsilon+2}\right)^{2}\,,\\
\Lambda_{\text{(e)a}} & =\frac{16\pi^{5}cR^{3}}{189}\left(\frac{k_{B}T_{(i)ex}}{\hbar c}\right)^{6}\text{Im}\left(\frac{\epsilon-1}{\epsilon+2}\right)\,,
\end{alignat}
where $\epsilon$ is the dielectric constant, $n_{V}$ the number
density of the gas inside the experimental box, and $r_{s}$ is the
sphere radius.

The damping factor $\Gamma_{\text{d}}$ in Eq.~\eqref{eq:fda2} has
been obtained by integrating the effect of the three decoherence sources.
Importantly, unlike in the case of non-inertial jitter, which is due
to the gas on the \emph{outside} and \emph{inside} the experimental
box, here only the gas environment \emph{inside }the experimental-box
decoheres the system. We finally remark that \emph{decoherence} channels,
unlike the technical noises discussed above, affect the density matrix
elements $\rho_{32}$ in Eq.~\eqref{eq:rho32}. Indeed, decoherence
presents a fundamental limitation to the degree of entanglement between
the two particles, which cannot be removed in a simple way using a
control experiment.

\subsection{Detectability of entanglement\label{subsec:Detectibility-of-entanglement}}

We consider the recently proposed optimized entanglement witness~\citep{chevalier2020witnessing}:
\begin{equation}
\mathcal{W}=\mathbb{I}\otimes\mathbb{I}-\sigma_{x}\otimes\sigma_{x}-\sigma_{y}\otimes\sigma_{z}-\sigma_{x}\otimes\sigma_{z}\,,\label{eq:witness}
\end{equation}
where $\sigma_{i}$ are the Pauli matrices. In particular, entanglement
is expected to be detected when: 
\begin{equation}
\langle\mathcal{W}\rangle=\text{Tr}(\mathcal{W}\rho)<0.\label{eq:witness-req}
\end{equation}
Using the density matrix elements in Eqs.~\eqref{eq:rho11}-\eqref{eq:rho32}
we eventually find 
\begin{align}
\langle\mathcal{W}\rangle= & 1-e^{-\Gamma_{\text{n}}/2-\Gamma_{\text{d}}/2}\left(\sin(\Delta\phi_{\uparrow\downarrow})+\sin(\Delta\phi_{\downarrow\uparrow})\right)\nonumber \\
+ & \frac{e^{-\Gamma_{\text{d}}}}{2}\left(e^{-2\Gamma_{\text{n}}}+\cos(\Delta\phi_{\downarrow\uparrow}-\Delta\phi_{\uparrow\downarrow})\right),\label{eq:witness no approx}
\end{align}
where $\Delta\Phi_{\downarrow\uparrow}$ ( $\Delta\Phi_{\uparrow\downarrow}$)
is given in Eq.~\eqref{eq:ud} (Eq.~\eqref{eq:du}).

If we consider the experimental values in Sec.~\ref{sec:setup} we
find that both $\Delta\Phi_{\downarrow\uparrow}$ and $\Delta\Phi_{\uparrow\downarrow}$
are small,$\sim0.01$, and hence the damping of coherences, $\Gamma_{\text{n}}$
and $\Gamma_{\text{d}}$, have to be even smaller, i.e. $\Gamma_{\text{n}},\Gamma_{\text{d}}\ll0.01$.
Hence we can further simplify Eq.~\eqref{eq:witness no approx} to
obtain a simple expression $\langle\mathcal{W}\rangle=\Gamma_{\text{n}}+\Gamma_{\text{d}}-\Phi_{\text{eff}},$
where $\Phi_{\text{eff}}$ is the effective entanglement phase given
in Eq.~\eqref{eq:phase}. The condition to witness entanglement is
thus $\Phi_{\text{eff}}>\Gamma_{\text{n}}+\Gamma_{\text{d}}$, which
in our specific case can be written as:

\begin{equation}
\Phi_{\text{eff}}>\Gamma_{\text{jitter}}+\Gamma_{\text{gg}}+\Gamma_{\text{d}},\label{eq:requirement-1}
\end{equation}
i.e. the effective entanglement phase must be larger than the damping
of the coherences.

Let us first consider the effect of non-inertial jitter while neglecting
other channels for the loss of visibility. The condition in Eq.~\eqref{eq:requirement-1}
reduces to $\Phi_{\text{eff}}>\Gamma_{\text{jitter}}$. We consider
the outside of experimental box\footnote{We consider the internal particle temperature $T_{\text{i}}=0.15\text{K}$,
which we assume to match the temperature inside the experimental box
$T_{\text{\text{e}}}=1\text{K}$, while the pressure inside the box
is $p_{\text{e}}=10^{-16}\text{Pa}$. As these values are substantially
lower than the corresponding values outside of the experimental box
we can sefely neglect their effect for non-inertial jitter.} to be at room temperature $T=300\text{K}$ and pressures from $p=10^{-12}\text{Pa}$
to $p=10^{-3}\text{Pa}$ (note that while this value of pressure is
still lower than the $p=10~\text{Pa}$ in current drop tower tubes
\citep{kulas2017miniaturized} so that their current microgravity
level is not sufficient for us, such pressures, called ultra-high-vacuum
(UHV) has already been achieved in very large volumes such as in gravitational
wave detectors and particle accelerators). In particular, one can
observe that for a reasonable mass and size of the box the condition
$\Phi_{\text{eff}}>\Gamma_{\text{jitter}}(p,T)$ is satisfied (see
Fig.~\ref{fig:boxsize}) -- lowering the outside pressure and temperature
would even further relax the constraints on the mass and size of the
experimental box. One can rewrite the constraint on the non-inertial jitter of the experimental
apparatus (in particular of the magnets) as a condition on the relative
acceleration noise $S_{AA}^{1/2}$. Specifically, we recall Eq.~\eqref{eq:GammaSimpleNIJ} which for
$t_{\text{a}}\ll t_{\text{e}}$ reduces to $\Gamma_{\text{jitter}}\sim S_{AA}m^{2}\Delta x^{2}t_{\text{e}}/\hbar^{2}.$
We further approximate the entanglement phase in Eqs.~\eqref{eq:ud},
\eqref{eq:du} and \eqref{eq:phase} as $\Delta\phi_{\text{eff}}\sim Gm^{2}t_{e}/(\hbar d).$
Finally, supposing $\Delta x\sim d$ we find that the condition $\Phi_{\text{eff}}>\Gamma_{\text{jitter}}$
reduces to $S_{AA}^{1/2}\apprle\sqrt{G\hbar/d^{3}}$ \citep{Andre}
which for $d\sim23\mu$m gives $S_{AA}^{1/2}\sim\text{1 fm\,\ensuremath{\text{s}^{-2}}}/\text{\ensuremath{\sqrt{\text{Hz}}}}$.  

We can also readily estimate the effect of GGN on the detectability
of entanglement, while neglecting other channels for the loss of visibility
-- specifically, the condition in Eq.~\eqref{eq:requirement-1}
reduces to $\Phi_{\text{eff}}>\Gamma_{\text{gg}}$. GGN is however
highly location dependent and will arise from atmospheric pressure
gradients, seismic activity and anthropogenic sources, among others
-- such noises cannot be measured directly by gravimeters which record
also non-gravitational contributions~\citep{harms2019terrestrial},
but have to be estimated from atmospheric and geophysical data as
well as modeling of anthropogenic activities. Here we will again exploit
Eq.~\eqref{eq:S2} which relates the GGN power spectral density (PSD)
to the more readily available acceleration noise PSD from the literature~\citep{saulson1984terrestrial,hughes1998seismic,thorne1999human}.
For the main GGN sources we will estimate the \emph{minimum} distance
from the experiment, $r_{\ensuremath{\text{min}}}$, which would still
allow the detection of entanglement.

Let us first estimate the GGN contribution arising from the main to
non-anthropogenic sources, namely from seismic and atmospheric activity.
In particular, we have the following acceleration PSDs \citep{saulson1984terrestrial,hughes1998seismic}:

\begin{alignat}{1}
S_{a_{\text{rand}}a_{\text{rand}}}^{\text{atmospheric}}(\omega)= & \frac{8\pi^{3}}{3}Gv_{s}^{2}\frac{\rho_{a}^{2}}{p_{a}^{2}}\frac{\vert\Delta p(\omega)\vert^{2}}{\omega^{2}},\\
S_{a_{\text{rand}}a_{\text{rand}}}^{\text{seismic}}(\omega)= & \frac{16\pi^{2}}{3}G\rho_{e}^{2}\vert\Delta X(\omega)\vert^{2},
\end{alignat}
for seismic and atmospheric sources, respectively. $v_{s}$ is the
speed of sound, $\rho_{a}$ ($p_{a}$) is air density (pressure),
$\Delta p$ is the pressure fluctuation, $\rho_{e}$ is the ground
density near the experiment, and $\Delta X$ is fluctuation of the
Earth's surface from the equilibrium position. By integrating over
all seismic and atmospheric mass movements as gravitational sources
of noise, one can estimate $\bar{a}\sim10^{-15}\text{ms}^{-2}/\sqrt{\text{Hz}}$
with $\alpha\sim4$ for $\omega/2\pi>10\text{Hz}$, and $\bar{a}\sim10^{-17}\text{ms}^{-2}/\sqrt{\text{Hz}}$
with $\alpha\sim0$ for $\omega/2\pi<10\text{Hz}$~\citep{saulson1984terrestrial,hughes1998seismic}.
Using Eq.~\eqref{eq:GammaggSimple} we find that the condition $\Phi_{\text{eff}}>\Gamma_{\text{gg}}$
is satisfied already if the bulk of the seismic and atmospheric GGN
originates at a characteristic distance $r_{\ensuremath{\text{min}}}\gtrsim10^{-2}\text{m}$
-- this indicates that GGN will likely not be a limiting factor even
for drop-tower experiment at the surface of the Earth, as these GGN
sources will be far more distant.

Human/anthropogenic movements can also contribute to the GGN -- for
example, a human walking near the experiment -- and one needs to
limit access to the experiment within a certain exclusion radius,
which we will again indicate with $r_{\ensuremath{\text{min}}}$.
We will consider two classes types of motion: a smooth continuous
straight-line motion, and discontinuous acceleration/deceleration
jerks. In particular, the acceleration noise PSD generated by an object
moving a constant velocity is given by~\citep{saulson1984terrestrial}:

\begin{alignat}{1}
S_{a_{\text{rand}}a_{\text{rand}}}^{\text{smooth}}(\omega)= & \frac{1}{\omega}\left(\frac{2Gm_{\text{ext}}}{b^{2}}\right)^{2}e^{-2\frac{b}{v_{\text{ext}}}\omega},\label{eq:smooth}
\end{alignat}
where $v_{\text{ext}}$ ($m_{\text{ext}}$) is the speed (mass), and
$b$ is the impact factor of the external object. Here we are only
interested to find an upper-bound on the GGN fluctuations and will
estimate the exclusion zone using the impact factor, i.e. $r_{\ensuremath{\text{min}}}\sim b$.
Although Eqs.~\eqref{eq:S2} and \eqref{eq:smooth} do not lead to
the simple formula in Eq.~\eqref{eq:GammaggSimple}, the GGN phase
fluctuations in Eq.~\eqref{eq:Gammagg} can be nonetheless readily
evaluated numerically. For concreteness, we consider a human of mass
$\sim100\text{kg}$ walking at a pace of $\sim1$m$s^{-1}$, and a
car of mass $\sim1000\text{kg}$ driving at $\sim10\text{m}\text{s}^{-1}$:
we find the exclusion zones $r_{\min}$$\sim2\text{m}$, and $r_{\min}\sim10\text{m}$,
respectively, which can be readily satisfied by restricting access
to the experimental building. GGN sources at larger distances do not
pose a limiting factor due to the favorable scaling of the GGN fluctuations
with the distance from the GGN source, i.e. $\Gamma_{\text{gg}}\sim e^{-2\frac{r_{\text{min}}}{v}\omega}/r_{\text{min}}^{6}$.
For example, for a plane of mass $\sim100\text{t}$ flying at speed
$\sim100$ms$^{-1}$ we find the exclusion radius $r_{\text{min}}\sim60\text{m}$.
Humans and cars can contribute to the GGN also in the non-adiabatic
regime of sudden acceleration/decelerations -- for example, during
regular weight transfers between steps. The latter effect can be characterized
by the following acceleration noise PSD~\citep{thorne1999human}:

\begin{alignat}{1}
S_{a_{\text{rand}}a_{\text{rand}}}^{\text{jerk}}(\omega)= & \frac{16G^{2}\Delta F_{\text{jerk}}^{2}}{P_{\text{gait}}\Delta t_{\text{jerk}}^{2}r_{\ensuremath{\text{min}}}^{6}\omega^{8}},\label{eq:jerk}
\end{alignat}
where $\Delta F_{\text{jerk}}$ is the change of the horizontal force
exerted by the human on the ground in a time interval $\Delta t_{\text{jerk}}$,
and $P_{\text{gait}}$ is the gait cycle of two steps. Following Ref.~\citep{thorne1999human}
we set $\Delta F_{\text{jerk}}\sim100N$, $\Delta t_{\text{jerk}}\sim20\text{ms}$
and consider a step time of $\sim400\text{ms}$ which results in the
frequency band $\frac{\omega}{2\pi}\sim[2.5\text{Hz},25\text{Hz}]$
and the exclusion radius $r_{\ensuremath{\text{min}}}\sim1\text{m}$.
We can also estimate the same effect for a car by assuming a stronger
change of force, say $\Delta F_{\text{jerk}}\sim10^{6}\text{Ns}^{-1}$,
which however only gives the exclusion radius $r_{\ensuremath{\text{min}}}\sim5\text{m}$.
Indeed, the walking style/pace or the type of car/vehicle will not
change drastically the exclusion radius: as can be seen from Eq.~\eqref{eq:S2}
and \eqref{eq:jerk}, the GGN phase fluctuations scale as $\Gamma_{\text{gg}}\sim r_{\text{min}}^{-8}$
quickly suppressing the effect of distant sources.

By conducting the experiment in an underground tunnel (still in a
free fall laboratory in the tunnel -- the tunnel acting as a drop
tower) the GGN can be further reduced. For example, at a depth of
$1$ km the typical distance $r_{\ensuremath{\text{min}}}$ from the
source (surface) increases to at least $\sim1$ km, resulting in a
significantly reduced GGN phase fluctuations $\Gamma_{\text{gg}}\apprle10^{-13}$.
In summary, we can conclude that in the QGEM experiment the effect
of GGN can be fully mitigated (see Fig.~\ref{fig:ggn}).

\section{Generality of analysis\label{sec:Generality-of-analysis}}

The analysis of the fluctuations leading to the depasing in Eqs.~\eqref{eq:GammaSimple}
and \eqref{eq:GammaggSimple} assumed that a force, $f_{\text{m}}$,
is used to create/recombine the superstitions. It is however instructive
to rewrite the force in terms of the transferred impulse

\begin{equation}
\Delta p\equiv f_{\text{m}}t_{a},
\end{equation}
where $t_{a}$ is the acceleration time-interval. This impulse $\Delta p$
is transferred four times during the creation/recombination parts
of the experiments (see Fig.~\ref{fig:scheme2}). In particular,
Eqs.~\eqref{eq:GammaSimple} and \eqref{eq:GammaggSimple} become

\begin{alignat}{1}
\Gamma_{\text{jitter}} & \approx\frac{16\gamma k_{B}T\Delta p^{2}}{\hbar^{2}M}\left[\frac{23}{15}t_{a}^{3}+t_{a}^{2}t_{e}\right],\label{eq:Gj1}\\
\Gamma_{\text{gg}} & \approx\frac{2\bar{a}^{2}\Delta p^{2}t_{a}^{2}}{\hbar^{2}}\left(\frac{d}{\bar{r}}\right)^{2}\left[\frac{C^{\alpha}(t_{a}+t_{e})^{2}t_{\text{exp}}^{\alpha-1}}{(2\pi){}^{\alpha}(\alpha-1)}\right],\label{eq:Gg1-1}
\end{alignat}
respectively. Interestingly, Eqs.~\eqref{eq:Gj1} and \eqref{eq:Gg1-1}
are now independent of the specific coupling between system and apparatus,
but depend only on generic experimental matterwave parameters. In
particular, Eqs.~\eqref{eq:Gj1} and \eqref{eq:Gg1-1} depend on
the experimental-box (mass $M$ and damping $\gamma$), the environment
outside the box (damping $\gamma$, temperature $T$, the strength
of the local acceleration fluctuations $\bar{a}$, a length-scale
parameter $\bar{r}$ characterizing the distance to the GGN sources,
a decay integer $\alpha>1$ which depends on the type of source, and
the constant $C=2\pi\times1\text{Hz}$), the geometry of the paths
(separation of the two interferometers $d$ ), the experimental times
(acceleration time-interval $t_{a}$, evolution time-interval $t_{e}$,
and total experimental time $t_{\text{exp}}$), and finally on the
transferred impulse, $\Delta p$. Moreover, as we have already discussed
in Sec. \ref{sec:noises} the noise fluctuations $\Gamma_{\text{jitter}}$,
$\Gamma_{\text{gg}}$ do not depend on the mass of the system, $m$.
The analysis of the technical noises considered here (non-inertial
jitter and GGN) is thus quite generic and could be adapted to any
matter-wave experiment.

\section{Discussion}

In this paper, we have shown that it is feasible to carry out an experiment
on quantum gravity induced entanglement of masses (QGEM) terrestrially
by going to a freely falling capsule. Of course, carrying out the
experiment in space will naturally form such a freely falling laboratory.
Under these circumstances, we have investigated the effect of non-inertial
(i.e., residual acceleration) and gravity gradient noise on the system
that still remains\footnote{There are also other specific systematic noises (i.e., technical noises)
depending on the specific mechanism of wavefunction splitting, some
of whose mitigations have already been extensively analysed \citep{marshman2020mesoscopic}).
Here we have taken conceptually the simplest route, namely exploiting
the \textit{equivalence principle} to get rid of the bulk of the gravitational
noise (all the acceleration noise) so that only \emph{relative} acceleration
noise due to non-inertial effects and the finite size of the experiment
remain. The other approach of measuring this purely classical noise
(see e.g. \citep{marshman2020mesoscopic,qvarfort2018gravimetry,armata2017quantum}),
and counteracting it by adjusting the detection in real-time or in
post-analysis will be discussed in a future paper.}. These types of noise, if untracked, induce an unknown \emph{relative}
acceleration between the interfering masses and the control and measuring
apparatus, which may appear as dephasing. We have thus carefully examined
the situations needed to keep the untracked parts below a threshold.

We have shown that non-inertial jitter (i.e., residual acceleration
noise) can only arise from non-gravitational effects, e.g. gas particles
and photons interacting with the experimental box which make it jiggle
about the geodesic motion (see Fig.~\ref{fig:conceptual}). We have
derived the loss of coherence due to non-inertial jitter from first
principles -- the noise originates mainly from the recoil of the
experimental-box due to collisions with dust particles. Thus non-inertial
jitter can be successfully mitigated by simply considering a heavy
experimental box (sometimes referred as capsule in the text) in a
low pressure/temperature environment. The heavier the experimental-box
the less it will recoil due to collisions with gas particles, and
the lower the pressure the lower the net recoil of the experimental-box
-- in both cases the residual Colella-Overhauser-Werner phase noise
can be strongly suppressed. Specifically, we have shown how different
pressure regimes constrain the mass/size of the experimental box to
successfully mitigate the corresponding loss of coherence~(see Fig.~\ref{fig:boxsize}).
For example, for pressures of $\sim10^{-6}$Pa outside the freely
falling capsule and at room temperature $\sim300$K, the non-inertial
component from random molecular kicks on a $\sim1\text{m}$ capsule
are low enough to enable a witnessing of the entanglement. Under the
above circumstances, for example, we are able to meet the acceleration
noise requirement pointed out in \citep{Andre} .

Furthermore, we have shown that the lowest order \emph{gravitational}
noise arises due to the finite size of the experiment and gravity
gradients (such a noise can be seen as the phase counterpart of the
tidal forces generated by external masses). We have considered the
main sources of gravitational noise: our estimates indicate that gravity
gradient noise from atmospheric and seismic sources is negligible,
while anthropogenic contributions can be fully mitigated by limiting
access to the immediate vicinity of the experiment (see Fig.~\ref{fig:ggn})
-- for example, to $\sim2m$ for humans, to $\sim10m$ for cars,
and to $\sim60\text{m}$ for planes. In summary, noise from gravitational
sources can be successfully mitigated by restricting access to the
experiment by placing in a dislocated building or possibly underground.

Finally, we have shown that \emph{relative} acceleration noise in
matterwave interferometry is intrinsically linked to momentum transfer,
$\Delta p$, between the system and experimental apparatus. We have
shown how the momentum transfer emerges in the \emph{full} interferometric
loop -- in particular, during the preparation and recombination of
the superposition when the system and experimental apparatus are coupled
(by magnetic fields or otherwise). Thus the dephasing effects discussed
in this work will become detrimental in \emph{any} matterwave interferometry
when significant forces/momentum transfers are used~\citep{brand2020bragg}
-- mitigation methods, such as the ones developed in this work, will
have to be adopted.

\section*{Acknowledgements }

MT and SB would like to acknowledge EPSRC grant No.EP/N031105/1, SB
the EPSRC grant EP/S000267/1, and MT funding by the Leverhulme Trust
(RPG-2020-197). AM's research is funded by the Netherlands Organisation
for Science and Research (NWO) grant number 680-91-119. RJM is funded
by a UCL Departmental Studentship. MSK was supported by the EPSRC
(EP/R044082/1) through the QuantERA ERA-NET Cofund in Quantum Technologies.

\bibliographystyle{unsrt}
\bibliography{accelerationnoise}

\end{document}